\documentclass[aps,prb,amsmath,twocolumn]{revtex4}
\usepackage{bm}
\usepackage{graphicx}
\begin{document}
\title{Minigap, Parity Effect and Persistent Currents in SNS Nanorings}

\author{Mikhail S. Kalenkov$^{1,2}$, Harald Kloos$^1$ and Andrei D. Zaikin$^{1,2}$}
\affiliation{$^1$Forschungszentrum Karlsruhe, Institut f\"ur Nanotechnologie,
76021, Karlsruhe, Germany\\
$^2$I.E. Tamm Department of Theoretical Physics, P.N.
Lebedev Physics Institute, 119991 Moscow, Russia}

\begin{abstract}
We have evaluated a proximity-induced minigap in the density of
states (DOS) of SNS junctions and SNS nanorings at an arbitrary
concentration of non-magnetic impurities. We have demonstrated
that an {\it isotropic} energy minigap in the electron spectrum
opens up already at arbitrarily weak disorder, while angle
resolved DOS at higher energies can remain strongly anisotropic.
The minigap value $\varepsilon_g$ can be tuned by passing a
supercurrent through an SNS junction or by applying a magnetic
flux $\Phi$ to an SNS ring. A non-monotonous dependence of
$\varepsilon_g$ on $\Phi$ has been found at weak disorder. We have
also studied persistent currents in isolated SNS nanorings. For
odd number of electrons in the ring we have found a non-trivial
current-phase (current-flux) relation which -- at relatively high
disorder -- may lead to a $\pi$-junction state and spontaneous
currents in the ground state of the system.
\end{abstract}


\maketitle

\section{Introduction}

In hybrid structures composed of a superconductor (S) and a normal
metal (N) Cooper pairs can penetrate into the latter thereby
significantly changing the properties of the system \cite{dG}. As
a result of this proximity effect the N-metal also acquires
superconducting properties being able to carry supercurrent and,
hence, exhibiting Josephson \cite{KI,many,Dubos} and Meissner
\cite{Z,BBS} effects.

Another interesting consequence of this proximity-induced
superconductivity is the existence of a minigap in the electron
spectrum of a normal metal. In the diffusive limit this minigap
$\varepsilon_g$ was found to be of order of the Thouless energy
$\varepsilon_{\rm Th}$ of this metal \cite{GK,BBS2,Been,BWBSZ}. In
strictly ballistic SN and SNS systems this gap is strongly
anisotropic \cite{Andreev}, and it vanishes for electrons
propagating parallel to the SN interface. For arbitrary
concentration of non-magnetic impurities the minigap dependence on
the electron elastic mean free path was studied in Ref.~\onlinecite{Pilgram}.
In SNS structures the proximity-induced
minigap $\varepsilon_g$ can be tuned by applying the phase
difference $\varphi$ across the N-metal \cite{KI,Zhou}.

Modifications of the normal metal density of states (DOS) due to
the proximity effect can be -- and were -- studied experimentally
with the aid of tunneling spectroscopy methods, see, e.g., Ref.
\onlinecite{Gue}. In this paper we point out that the
proximity-induced minigap in SN sandwiches can be directly
measured with the aid of the superconducting parity effect
\cite{AN92,Tuo92,SchZ94}. Indeed, the ground state energies of
isolated superconducting system with odd and even electron numbers
should differ exactly by the value of the gap in the electron
spectrum. In the case of superconducting grains it is just the BCS
gap, while in SN and SNS structures this value should be set by
the proximity-induced minigap $\varepsilon_g$. Hence, the latter
can be directly observed in an experimental setup similar to that
used, e.g., in Ref. \onlinecite{Laf93}.

\begin{figure}
\centerline{
\includegraphics[width=70mm]{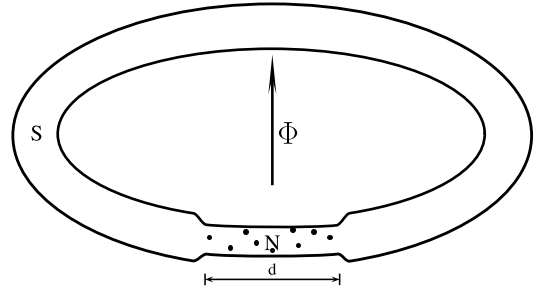}
}
\caption{A superconducting ring with embedded normal metal of length $d$. }
\label{snsring}
\end{figure}

The possibility of tuning the minigap value
$\varepsilon_g(\varphi)$ by passing the supercurrent through SNS
junctions provides additional ways to independently test
theoretical predictions. One can, for instance, consider an
isolated superconducting ring with an embedded layer of a normal
metal, the so-called SNS ring \cite{SZ}. This system is depicted
in Fig. 1. Applying an external magnetic flux $\Phi$ to such a
system one induces persistent currents (PC) circulating inside the
ring. Both the magnitude and the flux dependence of such currents
will depend on the parity of the total number of electrons in the
ring \cite{SZ,SZ2}. The difference between PC values for odd
($I_o$) and even ($I_e$) ensembles is related to the minigap value
$\varepsilon_g(\varphi )$. This relation acquires a particularly
simple form in the limit $T \to 0$ in which case one finds
\cite{SZ}
\begin{equation}
I_o(\Phi )=I_e(\Phi )+2e\dfrac{\partial \varepsilon_g(\varphi)}{\partial\varphi}.
\label{Ioe}
\end{equation}
The last term in this equation describes the contribution to the current from
the ``odd'' electron occupying the lowest available state above the minigap
$\varepsilon_g(\varphi )$ in the density of states of the normal metal.

For a broad range of system parameters the phase difference $\varphi$ across the SNS
junction is linked to the external magnetic flux $\Phi$ by means of the
standard relation $\varphi =2\pi \Phi /\Phi_0$ ($\Phi_0$ is the flux quantum)
which will also be assumed to hold throughout this paper. Eq. (\ref{Ioe}) can
be used, on one hand, for an independent study of the phase dependence of
the minigap $\varepsilon_g(\varphi)$ and, on the other hand, for further
investigations of the parity-affected persistent currents in SNS nanorings.

The structure of our paper is as follows. In Sec. II we will
employ the quasiclassical formalism of Eilenberger equations and
evaluate the phase-dependent minigap $\varepsilon_g(\varphi)$ in
SNS systems at an arbitrary concentration of non-magnetic
impurities. We will demonstrate that an {\it isotropic} minigap
opens up in the normal metal already in the limit of very weak
disorder. At sufficiently large values of the electron elastic
mean free paths we recover a {\it non-monotonous} dependence of
the minigap on $\varphi$. In Sec. III we will analyze an interplay
between parity effect and persistent currents in SNS nanorings
with an arbitrary impurity concentration. We will show that in the
case of odd total number of electrons in the ring the dependence
of PC on the applied magnetic flux acquires non-trivial features
which might lead to a $\pi$-{\it junction behavior} and {\it
spontaneous currents} in the ground state of the ring. A brief
discussion of our key observations is presented in Sec. IV.

\section{Phase-dependent minigap at arbitrary
impurity concentrations}

Let us analyze thermodynamic properties of SNS rings presented in
Fig.~\ref{snsring}. We will stick to the simplest case of a
superconductor  with singlet isotropic pairing and assume that
both NS interfaces are fully transparent. The length of the normal
metal layer and its cross section are denoted respectively as $d$
and ${\cal A}$.

Our analysis is based on the Eilenberger equations
\cite{Eil68,BWBSZ} for the energy-integrated retarded $2\times2$
matrix Green functions $\hat g$
\begin{gather}
        \left[
        \varepsilon \hat\tau_3-\hat\Delta(\bm{r})-
        \hat\Sigma(\bm{r}, \varepsilon),
        \hat g (\bm{p}_F,\bm{r}, \varepsilon)
        \right]=
        -i\bm{v}_F \nabla \hat g (\bm{p}_F,\bm{r}, \varepsilon),
        \label{quaseq}
        \\
        \hat g^2 (\bm{p}_F,\bm{r}, \varepsilon)=1,
\end{gather}
where $[\hat a,\hat b]=\hat a\hat b-\hat b\hat a$, $\varepsilon$
is the quasiparticle energy, $\bm{p}_F=m\bm{v}_F$ is the electron
Fermi momentum vector and $\hat\tau_3$ is the Pauli matrix. The
matrices  $\hat g$ and $\hat\Delta$ have the standard form
\begin{equation}
        \hat g =
        \begin{pmatrix}
                g & f \\
                f^+ & -g \\
        \end{pmatrix}, \quad
        \hat\Delta=
        \begin{pmatrix}
                0 & \Delta \\
                -\Delta^* & 0 \\
        \end{pmatrix},
\end{equation}
where $\Delta$ is the BCS superconducting order parameter chosen
to be spatially constant in the superconductor and equal to zero
in the normal metal. Electron scattering on non-magnetic
impurities is accounted for by the self-energy term $\hat\Sigma$.
Within the Born approximation one has
\begin{equation}
\hat\Sigma( \bm{r}, \varepsilon) = -i\frac{v_F}{2 \ell} \left< \hat g
(\bm{p}_F,\bm{r}, \varepsilon) \right>,
\label{se}
\end{equation}
where $\ell$ is the electron elastic mean free path and angular brackets $
\left< ... \right>$ denote averaging over the Fermi momentum directions.

The current density and the angle resolved local density of states are
defined as
\begin{gather}
\label{curr}
        \bm{j}(\bm{r})=
        eN_0 {\rm Re}
        \int\limits_{-\infty}^{\infty}
        d\varepsilon
        \tanh\left(\dfrac{\varepsilon}{2T}\right)
        \left< \bm{v}_F g(\bm{p}_F,\bm{r},\varepsilon)\right>,
        \\
        \nu(\bm{p}_F,\bm{r},\varepsilon)=
        {\rm Re} g(\bm{p}_F,\bm{r},\varepsilon),
\end{gather}
where DOS $\nu(\bm{p}_F,\bm{r},\varepsilon)$ is normalized to its
normal state value at the Fermi energy $N_0=mp_F/2\pi^2$. Let us
also note that the above quasiclassical formalism and, in
particular, the description of electron scattering in terms of the
self-energy (\ref{se}) are applicable provided the number of
conducting channels in the system is large $N_{Ch}={\cal A}
p_F^2/4\pi \gg 1$. This inequality will be assumed to hold
throughout the whole calculation \cite{GZ}. In addition, we will
assume that the length of the N-metal strongly exceeds the
superconducting coherence length $d \gg \xi_0\sim v_F/\Delta$.

Our first goal is to analyze the behavior of DOS
$\nu(\bm{p}_F,\bm{r},\varepsilon)$ in the normal metal as a
function of the phase difference $\varphi = 2\pi \Phi /\Phi_0$ for
arbitrary values of elastic mean free path $\ell$. To begin with,
let us recall that in a strictly ballistic limit $\ell \to \infty$
the spectrum of an SNS junction below the BCS energy gap is formed
by discrete Andreev levels with energies $E_n$. In the limit $E_n
\ll \Delta$ the values $E_n$ read \cite{Andreev, KI}
\begin{equation}
E_n = \frac{|v_x|}{2d} [\pi(2n+1)+ \varphi \mathrm{sgn}\, v_x],
\label{andlev}
\end{equation}
where $v_x$ is the $x$-component of the Fermi velocity vector
$\bm{v}_F$. This expression demonstrates that the proximity
induced gap in the spectrum of a ballistic SNS system is strongly
anisotropic. It depends on the direction of the Fermi velocity and
vanishes for electrons propagating parallel to NS interfaces. In
other words, the trajectories for such electrons do not cross NS
interfaces and, hence, these electrons do not suffer Andreev
reflection and do not ``feel'' any proximity effect.

\begin{figure}[t]
\centerline{
\includegraphics[scale=0.80]{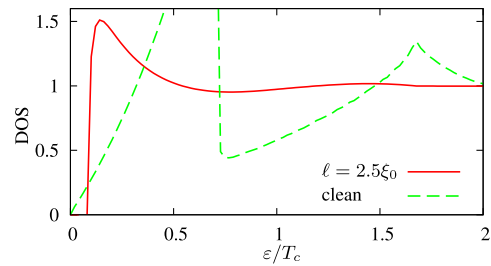}
}
\caption{Angle averaged DOS in the middle of the
normal layer. Here and below we choose $d= 10 \xi_0$. }
\label{DOS-phi000}
\end{figure}

The situation changes significantly in the presence of already
very weak disorder. In this case all electrons in the N-metal get
scattered by non-magnetic impurities and, sooner or later, their
trajectories hit one of the NS interfaces and electrons get
Andreev-reflected. As a result, an {\it isotropic}
proximity-induced minigap $\varepsilon_g(\varphi)$ develops in the
density of states of the normal metal. The value of this minigap
depends both on the electron mean free path $\ell$ and the phase
difference $\varphi$ but it remains independent of the momentum
direction for all values of $\ell$ and $\varphi$.

\begin{figure}[t]
\centerline{
\includegraphics[scale=0.80]{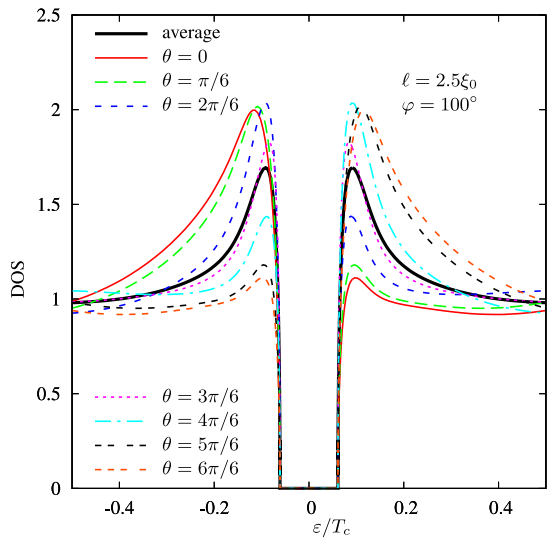}
}
\caption{The angle resolved DOS in the middle of
the normal layer (thin lines) together with the angle averaged DOS
(bold solid line). The angle between quasiparticle momentum and
the $x$-direction (normal to NS interfaces) is denoted as
$\theta$.  } \label{DOS-4-phi100-ang}
\end{figure}

This property is illustrated in Figs. \ref{DOS-phi000} and
~\ref{DOS-4-phi100-ang}. Fig. \ref{DOS-phi000} shows DOS inside
the normal metal with disorder together with that evaluated in the
ballistic limit $\ell \to \infty$. An important difference between
the two curves is observed at low energies: The proximity-induced
minigap is strictly zero in the ballistic limit while it opens up
in the presence of disorder. At higher energies electron
scattering on disorder broadens and eventually washes out peaks
corresponding to Andreev levels (\ref{andlev}).

Fig. ~\ref{DOS-4-phi100-ang} displays a typical energy dependence
of the angle resolved density of states $\nu(\bm{p}_F, \bm{r},
\varepsilon)$ inside the normal layer of an SNS structure in the
presence of disorder. We observe that $\nu(\bm{p}_F, \bm{r},
\varepsilon)$ vanishes for all $\varepsilon < \varepsilon_g (
\varphi)$ independently of the direction of the Fermi momentum.
Thus, already very weak disorder \cite{FN1} makes the minigap
isotropic for all values of $\varphi$. In addition, we have
verified that the minigap value does not depend on the coordinate
inside the normal metal.

At the same time at energies just above the minigap $\varepsilon >
\varepsilon_g$ the anisotropy in the density of states is clearly
observable for sufficiently large values of the electron mean free
path $\ell$, see Fig.~\ref{DOS-4-phi100-ang}. This anisotropy
decreases with decreasing mean free path $\ell$, and in the dirty
limit $\ell \lesssim \xi_0$ the density of states $\nu(\bm{p}_f,
\bm{r}, \varepsilon)$ becomes almost isotropic. Anisotropic
behavior of the density of states in the ultra-clean limit $\ell
\gg d$ can easily be understood if one observes that the minigap
value in the latter limit is of order $\varepsilon_g \sim v_F/\ell
\ll v_F/d$. Hence, for quasiparticles with energies just above the
minigap $v_F/d \gg \varepsilon \gtrsim v_F/\ell$ one can estimate
$v_x \sim v_F d/\ell \ll v_F$, i.e. these states are formed
predominantly by grazing electrons.

\begin{figure}[t]
\centerline{
\includegraphics[scale=0.80]{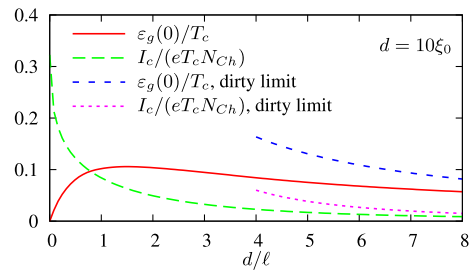}
}
\caption{The minigap $\varepsilon_g(0)$ and the zero temperature
Josephson critical current of an SNS junction as a function of $d/\ell$.
For sufficiently short values of $\ell$ both quantities approach the
corresponding results derived in the diffusive limit from the Usadel equation analysis.}
\label{je}
\end{figure}

Let us also note that in the normal metal at energies just above
the minigap $\varepsilon > \varepsilon_g$ the angle averaged DOS
$\nu(\bm{r}, \varepsilon) =\left<\nu(\bm{p}_F, \bm{r},
\varepsilon)\right>$ depends on energy as
\begin{equation}
\nu(\bm{r}, \varepsilon)\propto (\varepsilon
-\varepsilon_g)^{1/2}. \label{sqrt}
\end{equation}
This square-root dependence appears to be universal for all mean
free path values $\ell$. In order to verify this statement we have
plotted $\ln \nu(\varepsilon )$ versus logarithm of the parameter
$ (\varepsilon -\varepsilon_g)/\varepsilon_g$ for different values
of $\ell$ (not shown). For all impurity concentrations at energies
above the minigap our numerical data points collapse on straight
lines corresponding to the same slope 1/2. For very clean systems
the dependence (\ref{sqrt}) is observed within narrower energy
interval above the gap, and in the limit $\ell \to \infty$ this
interval shrinks to zero faster than the minigap itself. On the
contrary, for dirtier systems Eq. (\ref{sqrt}) applies for broader
energy intervals and match with the corresponding dependence
\cite{Zhou} established from Usadel equations in the diffusive
limit.

The dependence of the minigap $\varepsilon_g$ at $\varphi =0$
on the mean free path $\ell$ is depicted in Fig. \ref{je}. We have evaluated the minigap
at $d=10 \xi_0$ and $d=10^3\xi_0$ and
found a very good agreement with earlier numerical results of Pilgram {\it et
  al.} \cite{Pilgram} obtained for NS structures with the normal layer
thickness equal to $10^2\xi_0$. Our numerical results for
$\varepsilon_g\equiv \varepsilon_g(0)$ are rather well
approximated by a simple analytical formula
\begin{equation}
\varepsilon_g=\frac{v_F}{d}\frac{ay}{by^2+cy+1},
\label{app1}
\end{equation}
where $y=d/\ell$ and $a \approx 0.47$, $b \approx 0.45$, $c
\approx 1.45$. With decreasing $\ell$ the minigap first grows as
$\varepsilon_g \approx 0.47v_F/\ell$, reaches its maximum
$\varepsilon_g^{\rm max} \approx 0.17v_F/d$ at $\ell \approx 0.67
d$ and then decays for small $\ell$ as $\varepsilon_g \approx 3.12
\varepsilon_{\rm Th}=1.04v_F\ell/d^2$ approaching the results of
Belzig {\it et al.} \cite{BBS2,BWBSZ} derived from the Usadel
equations analysis. Here and below $\varepsilon_{\rm
Th}=v_F\ell/3d^2$ is the Thouless energy of a normal metal.

\begin{figure}[t]
\centerline{
\includegraphics[scale=0.80]{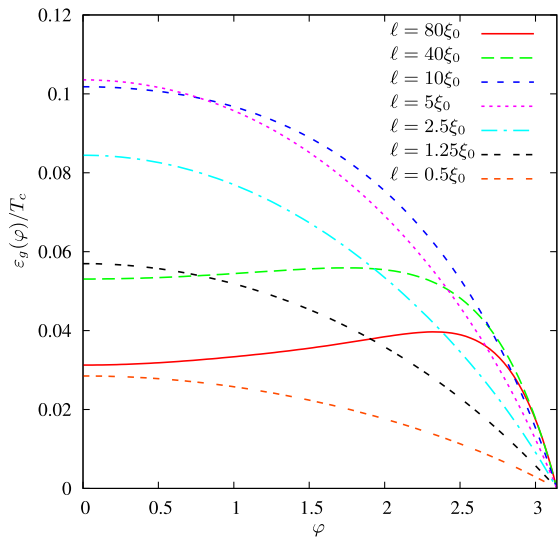}
}
\caption{Phase dependence of the minigap $\varepsilon_g(\varphi )$ for
an SNS structure at different values of the electron mean free path $l$.}
\label{minigap}
\end{figure}

Phase dependence of the minigap is shown in Fig.~\ref{minigap} for
the mean free path values ranging between quasi-ballistic ($\ell
\gg d$) and diffusive ($\ell \ll d$) regimes. A somewhat
unexpected non-monotonous dependence of the minigap value on the
phase difference is observed in the ultra-clean limit $\ell\gg d$.
We believe that this effect is caused by a non-trivial
redistribution in the angle resolved DOS in the presence of the
phase twist $\varphi$ across the N-layer. With decreasing the mean
free path the maximum of the function $\varepsilon_g(\varphi )$
moves away from the point $\varphi=\pi$ towards smaller values of
the phase difference. Eventually the maximum of
$\varepsilon_g(\varphi )$ reaches the point $\varphi =0$ and this
dependence becomes monotonous at all smaller values of $\ell$. In
the diffusive limit $\ell \ll d$ our results approach those
obtained by Zhou {\it et al.} \cite{Zhou} from Usadel equations,
cf. Fig. 2 in Ref. \onlinecite{Zhou}.

In the limit $\ell \ll d$ the minigap dependence on $\varphi$ is
rather well approximated by a simple parabola
\begin{equation}
\varepsilon_g (\varphi) \simeq \varepsilon_g (0)(1-\varphi^2/\pi^2).
\end{equation}
Note that although this dependence is in agreement with our
numerical results as well as with those displayed in Fig.~2 of
Ref. \onlinecite{Zhou}, it does not agree with Eq. (5) from that
paper at small phases $\varphi \ll \pi$.

\section{Persistent currents and parity effect}

We now go over to the calculation of persistent currents in
isolated SNS nanorings. As the total number of electrons in the
ring is fixed to be either even or odd it is necessary to account
for the parity effect. This task will be accomplished within the
parity projection technique \cite{JSA94,GZ94,AN94} recently
adapted \cite{SZ,SZ2} to the calculation of PC in superconducting
nanorings.

Let define and evaluate the even/odd parity projected
thermodynamic potentials
\begin{equation}
                \Omega_{e/o}=\Omega_f-T
                \ln\left[
                \dfrac{ 1\pm e^{-A}}{2}
                \right],
                \quad
                A=(\Omega_b - \Omega_f)/T,
                \label{Omega_eo}
\end{equation}
where $\Omega_f$ is the standard grand canonical thermodynamic
potential and $\Omega_b$ is obtained from $\Omega_f$ by expressing
the latter as a sum over the Fermi Matsubara frequencies
$\omega_f=\pi T(2n+1)$ with subsequent substitution of $\omega_f$
by the Bose frequencies $\omega_b=2\pi Tn$. For further details we
refer the reader to the paper \cite{GZ94}.

With the aid of the Eilenberger quasiclassical formalism \cite{Eil68} it is
easy to cast the grand canonical  thermodynamic potential of the system
$\Omega_f$ to the following form
\begin{equation}
                \Omega_f = \tilde \Omega -
                4N_0 T\int d\bm{r}
                \int\limits_0^{\infty}d\varepsilon
                \ln \left[2\cosh(\varepsilon/2T)\right]\nu(\bm{r}, \varepsilon),
                \label{omegaf}
\end{equation}
where $\tilde\Omega = \int d \bm{r} |\Delta^2(\bm{r})|/\lambda +
\sum_{ \bm{k}} \xi_{ \bm{k}}$, $\lambda$ is the BCS coupling
constant and $\xi_{ \bm{k}}$ is the single particle energy.
Similarly, for $\Omega_b$ we find
\begin{equation}
                \Omega_b = \tilde\Omega -
                4N_0 T\int d\bm{r}
                \int\limits_0^{\infty}d\varepsilon
                \ln \left[2\sinh(\varepsilon/2T)\right]\nu(\bm{r},
                \varepsilon).
                \label{omegab}
\end{equation}
With the aid of the above expressions it is now possible to directly evaluate
PC circulating in a superconducting ring both for even and odd electron
ensembles. Taking the derivative of $\Omega_{f/b}$ with respect to the phase
difference one finds \cite{SZ,SZ2}:
\begin{equation}
                I_{e/o}=I_f\pm \dfrac{I_b-I_f}{\exp(A) \pm 1}.
                \label{Ieo1}
\end{equation}
where the upper (lower) sign corresponds to the even (odd) ensemble and
\begin{equation}
                I_{e/o}=2e\dfrac{\partial \Omega_{e/o}}{\partial\varphi},
                \quad
                I_{f/b}=2e\dfrac{\partial \Omega_{f/b}}{\partial\varphi}.
\end{equation}

It is well known that the parity effect is mostly pronounced in
the low temperature limit in which case the parameter $A$
(\ref{Omega_eo}) is much larger than one \cite{GZ94} $A\gg 1$.
Evaluating the integrals in Eqs. \eqref{omegaf}, \eqref{omegab} at
temperatures $T \ll |\varepsilon_g(\varphi)|$ we obtain
\begin{equation}
 A(T)=A_N(T)+A_S(T),
\end{equation}
where
\begin{gather}
A_N(T)\sim N_0V_N T^{3/2}\varepsilon_g^{-1/2}(\varphi)\exp(-\varepsilon_g(\varphi)/T),
\\
A_S(T)\sim N_0V_ST^{1/2}\Delta^{1/2}\exp(-\Delta/T)
\label{A}
\end{gather}
are the contributions respectively from the normal and
superconducting parts of the ring with the corresponding volumes
$V_N$ and $V_S$. The term $A_N$ was evaluated making use of the
dependence (\ref{sqrt}) for DOS at energies $\varepsilon$ just
above the minigap.

Since throughout this paper we always assume the minigap to be
smaller than the superconducting gap, $\varepsilon_g (\varphi ) <
\Delta$, at sufficiently low $T$ the term $A_N$ dominates over
$A_S$ provided the volume ratio $V_N/V_S$ is not too small.
Assuming that the contribution $A_S$ can be neglected, from the
condition \cite{GZ94} $A(T^*) \sim 1$ we arrive at the estimate
for the crossover temperature $T^*$. above which the difference
between thermodynamic potentials for even and odd ensembles is
negligible and the parity effect is smeared out by thermal
fluctuations. Within the logarithmic accuracy we find
\begin{equation}
T^* \approx \varepsilon_g (\varphi )/\ln (N_0V_N\varepsilon_g
(\varphi )).
\end{equation}
In the whole temperature range $T <T^*$ where the parity effect
remains pronounced the temperature effect on the Josephson current
across SNS junction turns out to be negligible for all values of
the mean free paths $\ell$. This is guaranteed by the inequality
$T^* \ll \min (\varepsilon_{\rm Th},v_F/d)$ which is always
satisfied for generic systems. Hence, in order to study PC in the
presence of the parity effect it is sufficient to restrict our
calculation to the zero temperature limit $T=0$. In this case from
the above analysis one immediately arrives at Eq. (\ref{Ioe})
which establishes a direct relation between PC values in the even
and the odd superconducting ensembles.

\begin{figure}[t]
\centerline{
\includegraphics[scale=0.80]{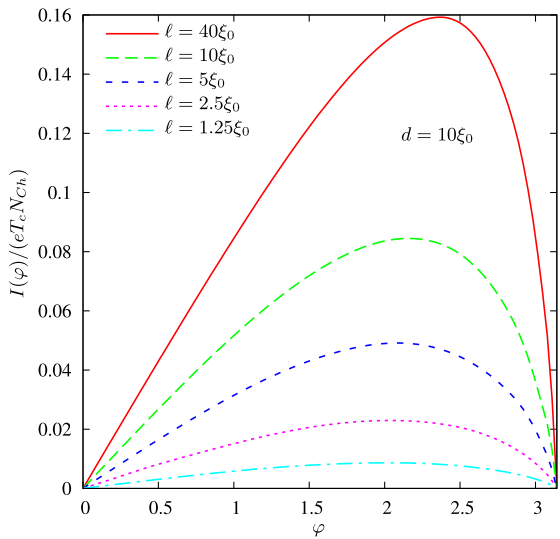}
}
\caption{Phase dependence for the zero
temperature Josephson current $I(\varphi )$ at different values of
the mean free path $l$.} \label{jcurrent}
\end{figure}

Let us first evaluate PC for the even ensemble $I_e$. At $T=0$ this current
identically coincides \cite{SZ,SZ2} with one calculated for the grand canonical
ensemble. The latter is easily evaluated with the aid of the quasiclassical
Eilenberger equations (\ref{quaseq}-\ref{curr}). The results for the current-phase relation
$I_e(\varphi )$ are displayed in Fig.~\ref{jcurrent} for various impurity concentrations.
The dependence of the critical current $I_C$ on the electron mean free path $\ell$ is
presented in Fig. \ref{je}.

\begin{figure}[t]
\centerline{
\includegraphics[scale=0.80]{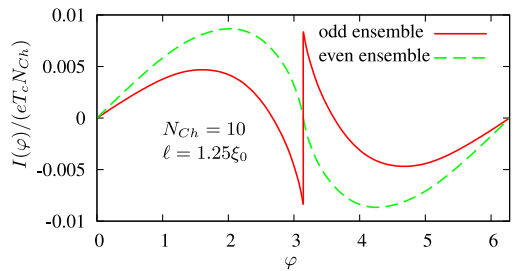}
}
\caption{Phase dependence of the Josephson current at $T=0$
for the odd and even number
of electrons in the ring.}
\label{jcurrent-N10}
\end{figure}

PC in the odd ensemble $I_o$ at $T=0$ can now be trivially evaluated by making
use of Eq. (\ref{Ioe}) and combining our results for $I_e(\varphi )$ with
those for the minigap $\varepsilon_g(\varphi )$ derived in the previous
section. The typical dependence $I_o(\varphi )$ is displayed in Fig.~\ref{jcurrent-N10}.
We observe that at sufficiently large values of $\varphi <\pi$ the
absolute value of the odd electron contribution to PC
$2e\partial \varepsilon_g/\partial\varphi$ exceeds
the term $I_e (\varphi )$ and the total current $I_o$ changes the sign. This
non-trivial parity-affected current-phase relation is specific for SNS rings
with disorder and it substantially differs from the current-phase relations
derived earlier for SNS rings with ballistic \cite{SZ} and resonant \cite{SZ2}
transmissions.

\begin{figure}[t]
\centerline{
\includegraphics[scale=0.80]{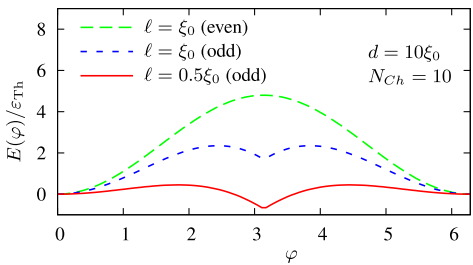}
}
\caption{Josephson energy $E(\varphi )$ of an SNS ring as a function of the
phase difference $\varphi$ for the even and odd ensembles.
The solid curve corresponds
to a $\pi$-junction state.}
\label{omega}
\end{figure}

At the same time, as in the previous cases \cite{SZ,SZ2}, in the
odd ensemble there emerges a possibility for a $\pi$-junction
state as well as for spontaneous currents in the ground state of
the system without any externally applied magnetic flux. In order
to analyze the situation it is sufficient to evaluate the ground
state energy of the SNS junction by integrating Eq. (\ref{Ioe})
with respect to the phase $\varphi$. One finds
\begin{equation}
E_o(\varphi )=E_e(\varphi )-\varepsilon_g(0)+\varepsilon_g (\varphi ),\;\;\;
E_e(\varphi )=\frac{1}{2e}\int\limits_{0}^{\varphi} I_e(\varphi )d\varphi ,
\end{equation}
where $E_{e/o}(\varphi )$ are the ground state energies of SNS junction for
even and odd number of electrons in the ring. While the energy $E_e(\varphi )$
is always non-negative and reaches its minimum at $\varphi =0$, in the odd
case the ground state energy $E_o(\varphi )$ can become negative reaching its
absolute minimum at $\varphi =\pi$. This physical situation of a
$\pi$-junction is illustrated in Fig. \ref{omega}.

It is easy to find out under which conditions the $\pi$-junction state becomes
possible. For that purpose it is sufficient to observe that for any impurity
concentration $E_e(\pi )=\alpha I_C/e$, where $I_C$ is the grand canonical
critical current at $T=0$ and $\alpha$ is a number of order one which depends
on the particular form of the current-phase relation. The $\pi$-junction
condition $E_o(\pi )<0$ is equivalent to the inequality
\begin{equation}
\varepsilon_g(0)>\alpha I_C/e.
\label{picond}
\end{equation}
From Fig.~\ref{je} it is obvious that in the many channel limit
the inequality (\ref{picond}) cannot be satisfied for sufficiently
large $\ell$, in which case $I_C$ is large and, on the contrary, the
minigap $\varepsilon_g(0)$ is small \cite{FN}. On the other hand,
for sufficiently short values of the mean free path $I_C \propto
\ell^2$ decays faster with decreasing $\ell$ as compared to the
minigap $\varepsilon_g(0) \propto \ell$, and the $\pi$-junction
state becomes possible. In particular, in the diffusive limit one
finds \cite{BWBSZ} $I_C \simeq 10.82\varepsilon_{\rm Th}/eR_N=1.53
eN_{Ch}v_F \ell^2/d^3$ and $\alpha \simeq 1.05$, where $R_N$ is
the Drude resistance of a normal metal.
Combining these results
with the expression for the minigap \cite{BBS2,Zhou}
$\varepsilon_g (0) \simeq 3.12\varepsilon_{\rm Th}$, from the
condition (\ref{picond}) we observe that in the odd case a
$\pi$-junction state is realized provided the number of conducting
channels in the junction $N_{Ch}$ is smaller than
\begin{equation}
N_{Ch} < 0.65 d/\ell. \label{condN}
\end{equation}
This condition is not very restrictive and it can certainly be
achieved in various experiments. For sufficiently dirty junctions
it allows for a formation of a $\pi$-junction state even in the
many channel limit. The condition (\ref{condN}) can also be
rewritten as
$$
g_N <1.73,
$$
where $g_N=R_q/R_N=8N_{Ch}l/3d$ is the dimensionless conductance
of a normal layer and $R_q=h/e^2 \approx 25.8$ k$\Omega$ is the
quantum resistance unit.

The condition for the presence of spontaneous currents in the
ground state of SNS rings with an odd number of electrons is
established analogously, one should only take into account an
additional energy of the magnetic field produced by PC circulating
inside the ring. The ground state with spontaneous currents is
possible provided the total energy of the ring $E_{\rm tot}(\pi)$
becomes negative, i.e.
\begin{equation}
        E_{\rm tot}(\pi)= 1.8\varepsilon_{\rm Th}
        \left[g_N-1.73\right]+\dfrac{(\Phi_0/2)^2}{2\cal L}< 0,
\end{equation}
where ${\cal L}$ is the ring inductance. This condition is more
stringent than that for the $\pi$-junction state, but can also be
satisfied provided the ring inductance ${\cal L}$ exceeds a
certain threshold value which, strictly speaking, depends on $g_N$
and can roughly be estimated as $\sim 0.1
\Phi_0^2/\varepsilon_{\rm Th}$.

\section{Discussion}

For clarity, let us briefly summarize our key observations. Our
analysis was focused on the two main issues, the proximity-induced
minigap in NS and SNS structures at an arbitrary concentration of
non-magnetic impurities and the effect of parity number on
persistent currents in SNS nanorings. We have demonstrated that
already weak disorder qualitatively modifies the density of states
in the normal metal of NS and SNS proximity structures. A striking
observation at this point is that the proximity-induced minigap
$\varepsilon_g$ turns out to be {\it isotropic} for an arbitrarily
weak disorder \cite{FN2}, even though DOS at energies above the
gap $\varepsilon
>\varepsilon_g$ may remain highly anisotropic, cf., e.g., Fig.
~\ref{DOS-4-phi100-ang}.  Another interesting observation is the
possibility of a non-monotonous dependence of the minigap
$\varepsilon_g$ on the applied magnetic flux $\Phi$ (or phase
$\varphi$) in SNS rings, cf. Fig.  ~\ref{minigap}.  Such a
non-monotonous dependence can be realized only in the limit of
weak disorder, whereas for not too long elastic mean free paths a
monotonous decrease of the minigap from its maximum value at $\Phi
=0$ to zero at $\Phi=\Phi_0/2$ was found.

Piercing an SNS ring by an external magnetic flux one induces
circulating persistent currents in such a ring. In an isolated
ring both the amplitude and flux dependence of such PC may
strongly depend on the electron parity number. Provided the number
of electrons in the ring is odd, one electron remains unpaired
down to $T=0$ and occupies the lowest available state above the
proximity-induced minigap $\varepsilon_g (\varphi )$. This
electron produces a countercurrent which -- in the limit of
relatively short electron mean free paths -- may significantly
modify the current-phase relation and yield a $\pi$-junction
behavior and spontaneous currents in the ground state of an SNS
ring flowing in the absence of any external magnetic flux. Note,
that although these observations look qualitatively similar to
earlier results derived for SNS rings with ballistic \cite{SZ} and
resonant \cite{SZ2} transmission, there also exist important
differences. In particular, the current-phase relation is entirely
different in the diffusive limit considered here. Also the
restriction on the number of conducting channels $N_{Ch}$ in the
normal metal (\ref{condN}) is less stringent that that formulated
in Refs. \onlinecite{SZ,SZ2}. This feature of diffusive SNS rings
is rather advantageous for possible experimental observation of
the effects discussed here.

Our results demonstrate that superconducting parity effect can be used in
order to directly measure both the magnitude and the flux dependence of
the minigap in SNS nanorings. This can be done, e.g., with the aid of a
setup similar to one used in Ref. \onlinecite{Laf93}. Such type of
measurements is not restricted by the number of channels in the normal
part of the ring and may serve as an alternative to tunneling spectroscopy
of the minigap. Superconducting rings with embedded carbon nanotubes might
be promising candidates in order to experimentally investigate parity-affected
persistent currents in SNS nanorings. Several groups \cite{Delft,L1,L2,Hels}
have recently reported observations of dc Josephson current in superconducting
junctions with carbon nanotubes. Therefore, it appears feasible to fabricate
and experimentally investigate SNS rings with carbon nanotubes which should
exhibit properties predicted and analyzed in our paper.

\centerline{\bf Acknowledgments}

\vspace{0.5cm}

We are grateful to D.S. Golubev, A.A. Golubov, M.Yu. Kupriyanov
and S.V. Sharov for useful discussions. This work is supported by
the European Community's Framework Programme NMP4-CT-2003-505457
ULTRA-1D "Experimental and theoretical investigation of electron
transport in ultra-narrow 1-dimensional nanostructures".


\begin{thebibliography}{}
\bibitem{dG} See, e.g., P.G. de Gennes, {\it Superconductivity of Metals and
    Alloys} (Benjamin, New York, 1966).
\bibitem{KI} I.O. Kulik, Zh. Eksp. Theor. Fiz. {\bf 57},
1745 (1969) [Sov. Phys. JETP {\bf 30}, 944 (1970)]; C. Ishii,
Progr. Theor. Phys. {\bf 44}, 1525 (1970).
\bibitem{many} K.K. Likharev, Sov. Tech. Phys. Lett. {\bf 2}, 12 (1976);
A.D. Zaikin and G.F. Zharkov, Fiz. Nizk. Temp. {\bf 7}, 375
  (1981) [Sov. J. Low Temp. Phys. {\bf 7}, 181 (1981)]; F.K. Wilhelm,
  A.D. Zaikin and G. Sch\"on, J. Low Temp. Phys. {\bf 106}, 305 (1997).
\bibitem{Dubos} P. Dubos, H. Courtois, B. Pannetier, F.K. Wilhelm,
   A.D. Zaikin, and G. Sch\"on, Phys. Rev. B {\bf 63},  064502 (2001).
\bibitem{Z} A.D. Zaikin. Solid State Commun. {\bf 41}, 533 (1982).
\bibitem{BBS} W. Belzig, C. Bruder, and G. Sch\"on, Phys. Rev. B {\bf 53},
  5727 (1996).
\bibitem{GK} A.A. Golubov and M.Yu. Kupriyanov, J. Low Temp. Phys. {\bf 70}, 83 (1988).
\bibitem{BBS2} W. Belzig, C. Bruder, and G. Sch\"on, Phys. Rev. B {\bf 54},
  9443 (1996).
\bibitem{Been} K.M. Frahm, P.W. Brouwer, J.A. Melsen, and C.W.J. Beenakker,
Phys. Rev. Lett. {\bf 76}, 2981 (1996).
\bibitem{BWBSZ} W. Belzig, F.K. Wilhelm, C. Bruder, G. Sch\"on and A.D. Zaikin,
Superlatt. Microstruct. {\bf 25}, 1251 (1999).
\bibitem{Andreev} A.F. Andreev, Zh. Eksp. Theor. Fiz. {\bf 49},
655 (1965) [Sov. Phys. JETP {\bf 22}, 455 (1966)].
\bibitem{Pilgram} S. Pilgram, W. Belzig, and C. Bruder, Phys. Rev. B {\bf 62},
  12462 (2000).
\bibitem{Zhou} F. Zhou, P. Charlat, B. Spivak, and B. Pannetier, J. Low
Temp. Phys. {\bf 110}, 841 (1998).
\bibitem{Gue} S. Gueron, H. Pothier, N.O. Birge, D. Esteve, and M.H. Devoret,
Phys. Rev. Lett. {\bf 77}, 3025 (1996).
\bibitem{AN92} D.V. Averin and Yu.V. Nazarov, Phys. Rev. Lett. {\bf 69}, 1993
(1992).
\bibitem{Tuo92} M.T. Tuominen, J.M. Hergenrother, T.S. Tighe, and M. Tinkham,
 Phys. Rev. Lett. {\bf 69}, 1997 (1992).
\bibitem{SchZ94} G. Sch\"on and A.D. Zaikin, Europhys. Lett., {\bf 26}, 695 (1994).
\bibitem{Laf93} P. Lafarge, P. Joyez, D. Esteve, C. Urbina, and M.H. Devoret,
Phys. Rev. Lett. {\bf 70}, 994 (1993).
\bibitem{SZ} S.V. Sharov and A.D. Zaikin, Phys. Rev. B {\bf 71}, 014518
  (2005).
\bibitem{SZ2} S.V. Sharov and A.D. Zaikin, Physica E {\bf 29}, 360 (2005).
\bibitem{Eil68} G. Eilenberger, Z. Phys. \textbf{214}, 195 (1968).
\bibitem{GZ} Generalization of the analysis to nanostructures with few
conducting channels and impurity scattering requires, in general,
going beyond the quasiclassical Eilenberger formalism, see A.V. Galaktionov
and A.D. Zaikin, Phys. Rev. B {\bf 65}, 184507 (2002).
\bibitem{FN1} The maximum value of the electron mean free path used in our
  numerical calculation was $l=80 \xi_0$.
\bibitem{JSA94} B. Janko, A. Smith, and V. Ambegaokar,
Phys. Rev. B {\bf 50}, 1152 (1994).
\bibitem{GZ94} D.S. Golubev and A.D. Zaikin,
Phys. Lett. A {\bf 195}, 380 (1994).
\bibitem{AN94} D.V. Averin and Yu.V. Nazarov, Physica B {\bf 203}, 310 (1994).
\bibitem{FN} It is worth stressing again that in the quasi-ballistic the
  minigap is small only in the many channel limit, while for SNS juctions
with few conducting channels the electron motion in the
transversal direction is quantized and the minigap remains
non-zero also in the ballistic limit. In this respect the
situation considered here differs drastically from that addressed
in Refs. \onlinecite{SZ,SZ2}.
\bibitem{FN2} Effective isotropization of some other quantities in hybrid proximity structures
can also be expected already in the weak disorder limit. For
instance, an angle-independent decay length of the Cooper pair
amplitude in SFS (superconductor-ferromagnet-superconductor)
structures with disorder has been found, see D.Yu. Gusakova, M.Yu.
Kupriyanov, and A.A. Golubov, cond-mat/0605137.
\bibitem{Delft} P. Jarillo-Herrero, J.A. van Dam, and L.P. Kouwenhoven,
Nature, {\bf 439}, 953 (2006).
\bibitem{L1} H.I. Jorgensen, K. Grove-Rasmussen, T. Novotny, K. Flensberg, and
P.E. Lindelof, cond-mat/0510200.
\bibitem{L2}  K. Grove-Rasmussen, H.I. Jorgensen, and P.E. Lindelof,
  cond-mat/0601371.
\bibitem{Hels} P. Hakonen, private communication.
\end{thebibliography}
\end{document}